\documentclass[%
 reprint,
 amsmath,amssymb,
 aps,
pre,
floatfix,
]{revtex4-2}

\usepackage{amsmath}
\usepackage{graphicx}
\usepackage{dcolumn}
\usepackage{bm}
\usepackage{xcolor}

\usepackage{caption}
\usepackage{subcaption}

\begin{document}

\title{Effects of Structural Inhomogeneity on Equilibration Processes in Langevin Dynamics}

\author{Omid Mozafar}
\email{omozafar@uwo.ca}
\affiliation{
Applied Mathematics Department, The University of Western Ontario, London, Ontario, Canada N6A 5B7
}
\author{Colin Denniston}
\email{cdennist@uwo.ca}
\affiliation{Physics and Astronomy Department, The University of Western Ontario, London, Ontario, Canada, N6A 3K7}

\begin{abstract}
In recent decades, computer experiments have led to an accurate and fundamental understanding of atomic and molecular mechanisms in fluids, such as different kinds of relaxation processes toward steady physical states. In this paper, we investigate how exactly the configuration of initial states in a molecular-dynamics simulation can affect the rates of decay toward equilibrium for the widely-known Langevin canonical ensemble. For this purpose, we derive an original expression relating the system relaxation time $\tau_{sys}$ and the radial distribution function $g(r)$ in the near-zero and high-density limit. We found that for an initial state which is slightly marginally inhomogeneous in the number density of atoms, the system relaxation time $\tau_{sys}$ is much longer than that for the homogeneous case and an increasing function of the Langevin coupling constant, $\gamma$. We also found during structural equilibration, $g(r)$ at large distances approaches 1 from \textit{above} for the inhomogeneous case and from \textit{below} for the macroscopically homogeneous one. \textbf{Keywords}: System Relaxation Time,  Molecular-Dynamics Simulation, Langevin Thermostat, Wallace's Entropy Expansion, Inhomogeneous Structure, Radial Distribution Function.
\end{abstract}

                            
\maketitle


\section{\label{intro}introduction}

One of the fundamental physical processes in the world is the relaxation process of many-body systems from any perturbation. Generally speaking, once a system relaxes, it becomes temporally invariant and hence one would define reliable and measurable quantities indicating the systems' properties. The word \textit{relaxation} was originally utilized by the Scottish physicist, James C. Maxwell, in 1867 to describe molecular processes \cite{Maxwell67}. Since then, many researchers have put significant effort into generalizing and explaining the basic concepts behind relaxation processes to be applicable to a wider range of phenomena \cite{Robinson, Jortner, abragam1983principles,Karmakar3675, Kevin96, gujrati2018,samanta2015, Torre2004}. In particular, in 1946, the Russian physicist, Yacov Frenkel, shed some light on the relaxation processes in liquids \cite{Frenkel46}, leading to locating the Frenkel Line in supercritical fluids \cite{Santoro2008, Brazhkin_2012, Trachenko_2015, Fomin_2016, Ghosh2018}. 

Over the past few decades, the invention of novel methods of conducting computer-based experiments with the advancement of algorithms and/or computers has opened a new chapter in the study of different kinds of relaxation processes \cite{Allen2017,leimkuhler2015molecular, Gao2016}. However, computer experiments sometimes produce nonphysical results due to imperfect algorithms or models, especially those related to thermostats \cite{Moosavi18, Davidchack2010,Rosta2009,Basconi2013}. In this work, by studying the temporal evolution of the radial distribution function in molecular-dynamics (MD) simulations \cite{rapaport_2004, kuksin2005,Dune2016, KOMATSU2004345} we show possible impacts of initial structural inhomogeneity on the relaxation processes in Langevin dynamics.

Measurement of the radial distribution function (RDF) in MD simulations (with a constant number of atoms) has a long history \cite{Yarnell73, Averbach73, Gruebel67, Kacner79, Fichthorn91}. Early efforts suffered greatly from the inherently small system sizes accessible to molecular dynamics. This spawned numerous works on accounting for all kinds of finite-size effects when measuring the RDF \cite{Kodama2006,Raman99,Dawass2018,Villamaina_2014,Kim2000}. Recent advances in computer technology now allows the study of larger and larger systems, i.e., $N>>1$. An unappreciated effect of this is the much longer relaxation times needed for \textit{long-wavelength} structural degrees of freedom. We study these kinds of effects in this paper.

This paper is organized as follows: Sec. \ref{background} reviews some of the physical concepts; Sec. \ref{theory} derives an original expression connecting the RDF and $\tau_{sys}$; Sec. \ref{methodology} provides details on the computer simulations and system features; Sec. \ref{result} presents and discusses the results obtained from our MD simulations; Sec. \ref{conclusion} summarizes the paper along with some conclusions and suggestions.

\section{\label{background}Background}

The radial distribution function is the normalized local density distribution within the system when one looks radially outwards from any particle. It can also be thought of as a measure of the probability of finding one particle of any shape and orientation located at a distance of $r$ from the center of mass (c.m.) of a specified reference particle. The RDF is of great significance in condensed matter physics as it can directly be related to the static structure factor, $S(k)$, and hence, determined experimentally from radiation scattering experiments, such as those using x rays and neutrons \cite{powderdiff2008}.  For an infinite isotropic and homogeneous system, the relation between the RDF and $S(k)$ is given by \cite{barrat2003,zhang2016, Strum93}
\begin{align}
  S(k) =1+4\pi\rho\int_0^{\infty} r^2[g(r)-1]\frac{\sin{kr}}{kr}dr,
  \label{eq_strucfactor}
\end{align}
where $\rho$ is the system averaged number density of atoms. Physically, $S(k)$ describes the system density response at wavelength $2\pi/k$ to a  weak enough, external perturbation \cite{Hansen2013}. Note that $S(k)$ is always nonnegative in equilibrium. 

The RDF is also used to link microscopic structural details to macroscopic properties under the Kirkwood-Buff (KB) solution theory \cite{KBtheory94}.  In the canonical ensemble, for example, the potential of mean force ($w$) between any pair of particles in the fluid is related to the RDF via \cite{chandler87}
\begin{align}
w_N(r)=-k_BT\ln{g_N(r)},
\label{eq_pmf}
\end{align}
where the subscript $N$ is added to highlight that the total number of particles is constant, and $T$ is the equilibrium temperature. The $w$ is conveniently written as a sum of two terms if the total potential energy is approximated by a sum of identical, independent pair potential energies, $u(r)$:
\begin{align}
w_N(r)=u(r)+\delta F_N(r),
\label{eq_pmf2}
\end{align}
where $\delta F_N(r)$ includes the effects of the solvent and is the canonical-ensemble average change of $F_N$, the Helmholtz free energy, of the fluid introduced by bringing two atoms from infinity to a (finite) distance $r$. If the process is done \textit{adiabatically}, $\delta F_N(r)$ can be considered as the work done on the solvent, that is the remaining $N-2$ atoms, during the process. Generally speaking, $\delta F_N$ is nonzero at finite number densities.  Using finite-size corrections \cite{Lebwitz61,hill56, Sal96, Heidari, Roman97, SALACUSE20083073}, it can be shown for $N>>1$ and $r$ greater than the correlation length $\xi$ at \textit{any} density, or all $r$ at $\rho\to0^+$,
\begin{align}
\delta F_N(r)= \frac{k_BT\chi^{\infty}_T}{N},
\label{eq_limitpmf}
\end{align}
where $\chi_T^{\infty}=S(0)$ is the reduced isothermal compressibility of an open system [in the thermodynamic limit (TL), that is $N,L\rightarrow\infty$, while $\rho$ is constant], which shares the same equilibrium state  \cite{Heidari}.

There is a great number of research articles in the literature attempting to determine short- or long-range behaviours of the RDF in order to understand a wider range of phenomena, such as the wetting phenomenon \cite{Henderson94, Hughes2014}. It has rigorously been shown for a system in equilibrium with an interparticle potential which either decays faster than a power law or is truncated at a finite cutoff radius, the RDF for $r>r_c$, where $0<r_c<\xi$ is the cutoff radius or the (effective) range of the pair potential energy, $u(r)$, is given by (in the TL) \cite{Dijkstra2000,Stopper2019,Vega1995, Savenko2005}
\begin{align}
g(r)=1+\left(\frac{ R_{\xi}\cos{\beta_\xi r}}{2\pi r}\right)e^{-r/\xi},
\label{eq5}
\end{align}
where $R_{\xi}$ is the residue of the Fourier transform of $[g(r)-1]$ corresponding to the pole(s) with the smallest positive imaginary part, and $\beta_\xi$ is some (real) constant. Equation (\ref{eq5}) shows $r[g(r)-1]$ in equilibrium  decays asymptotically to zero exponentially, either monotonically (with $\beta_\xi=0$) or sinusoidally (with $\beta_\xi\neq0$) \cite{Vorontsov2008,Evans2009, montero2017}.

In the canonical ensemble, the average entropy $\langle S_N\rangle$ of an atomic fluid at a temperature $T$ with $N$ indistinguishable atoms, described by the canonical space and momentum coordinates $(\textbf{r}_1,\textbf{p}_1),\dots,(\textbf{r}_N,\textbf{p}_N)$, can be calculated  from the generalized Gibbs-Shannon formula \cite{Plischke2006,LANDAU1980}:
\begin{align}
\begin{split}
 \langle S_N\rangle=-\frac{k_B}{N!}\int d\textbf{r}_1d\textbf{p}_1\dots d\textbf{r}_Nd\textbf{p}_N \mathcal{P}_N\ln{[h^{3N}\mathcal{P}_N]},   
\end{split}
\label{eq_entropy}
\end{align}
where $\mathcal{P}_N$ is the (unnormalized) phase-space probability density, $k_B$ is the Boltzmann's constant, $h$ is the Planck's constant, and  $N!$ accounts for indistinguishablity of particles. If $H_N$, the total Hamiltonian, separates into additive terms for the potential and kinetic energy, the phase-space probability density can be factorized as
\begin{align}
\mathcal{P}_N= g_N(\textbf{r}_1,\dots,\textbf{r}_N)\prod_{i=1}^{N}f_1(\textbf{p}_i),
\label{eq_prob2}
\end{align}
where $f_1(\textbf{p}_i)$ is the $1$-particle momentum probability density:
\begin{align}
f_1(\textbf{p}_i)=\rho(2\pi mk_BT)^{-3/2}e^{-|\textbf{p}_i|^2/2mk_BT}
\label{eq_proboneparticle}
\end{align} 
Equation ($\ref{eq_prob2}$) serves as a definition for the $N$-particle positional distribution function, $g_N(\textbf{r}_1,\dots,\textbf{r}_N)$; Physically, it is a measure of the joint probability of finding the particle $1$ at position $\textbf{r}_1,\dots$, and particle $N$ at position $\textbf{r}_N$. Utilizing the generalized Kirkwood superposition approximation, one can factorize $g_N(\textbf{r}_1,\dots,\textbf{r}_N)$ as \cite{green52,Kir42}
\begin{align}
\begin{split}
 g_N(\textbf{r}_1,\dots,\textbf{r}_N)=g_N(\textbf{r}_1,\textbf{r}_2)\times\dots\times g_N(\textbf{r}_{N-1}, \textbf{r}_N)\\\times \delta g_N(\textbf{r}_1,\textbf{r}_2,\textbf{r}_3)\times\dots\times \delta g_N(\textbf{r}_1,\dots,\textbf{r}_N),    
\end{split}
\label{eq_Ndis}
\end{align} 
where, for example,
\begin{align*}
 \delta g_N(\textbf{r}_1,\textbf{r}_2.\textbf{r}_3)\equiv \frac{g_N(\textbf{r}_1,\textbf{r}_2,\textbf{r}_3)}{g_N(\textbf{r}_1,\textbf{r}_2)g_N(\textbf{r}_1,\textbf{r}_2)g_N(\textbf{r}_2,\textbf{r}_3)}.
\end{align*}
By substituting Eq.($\ref{eq_prob2}$) into Eq.($\ref{eq_entropy}$) and using Eqs.(\ref{eq_proboneparticle}) and (\ref{eq_Ndis}), an entropy expansion is obtained, which is well approximated by  \cite{Wallace1987,Laird1992,Widom2019,Baranyai1990,BORZSAK1992227, Baranyai1989}
\begin{align}
\langle S_N\rangle=\langle S_N^{id}\rangle+S_N^{(2)}  
\label{eq13}
\end{align}
for high (near the freezing point) and near-zero densities, where the first term on the r.h.s is the ideal-gas mean entropy, and the second term is the $2$-body (excess) entropy.

For a homogeneous and isotropic fluid, one can express the 2-body (excess) entropy in terms of the RDF \cite{Gallo2015} and $w$ [by using Eq.(\ref{eq_pmf})]:
\begin{align}
S_N^{(2)}=\frac{ N\rho}{2T}\int d\mathbf{r}g_N(r)w_N(r)-\frac{Nk_B}{2}
\label{eq_2body}
\end{align}
Substituting $w_N(r)$ in Eq.(\ref{eq_2body}) with the expression in Eq.(\ref{eq_pmf2}), the following expression is obtained: 
\begin{align}
S_N^{(2)}=\frac{\langle U_N\rangle}{T}+\delta S_N^{(2)}-\frac{Nk_B}{2},
\label{eq_2bodynew}
\end{align}
where $\langle U_N\rangle$ is the system's total potential energy, $Nk_B/2$ is the entropy contribution of the canonical ensemble, and
\begin{align}
\delta S_N^{(2)}=\frac{N\rho}{2T}\int d\mathbf{r}g_N(r)\delta F_N(r),
\label{eq_2bodynew1}
\end{align}
which is the solvent entropy contribution and referred to as the residual 2-body (excess) entropy. For low densities, one can show for $N>>1$,
\begin{align}
\lim_{\:\rho\:\to\:0^+}\delta S_N^{(2)}= \frac{Nk_B\chi_T^{\infty}}{2}
\label{eq_limitresidual}
\end{align}
by direct substitution of Eq.(\ref{eq_limitpmf}) into Eq.(\ref{eq_2bodynew1}). Utilizing Eqs.(\ref{eq_2bodynew}) and (\ref{eq_limitresidual}), the near-zero density limit of Eq.(\ref{eq13}) can be written for $N>>1$ as
\begin{align}
\lim_{\:\rho\:\to\:0^+}\langle S_N\rangle=\langle S_N^{id}\rangle+\frac{\langle U_N\rangle}{T}+\frac{Nk_B}{2}(\chi_T^{\infty}-1).  
\label{eq14}
\end{align}
In the ideal-gas limit, $\langle U_N\rangle=0$ and $\chi_T^{\infty}=1$.

\section{\label{theory} Theory}

A closed system is in thermodynamic equilibrium, provided that its Helmholtz free energy is \textit{globally} minimized \cite{reif2009f, greiner1995}. The Helmholtz free energy for a system in contact with a heat bath is mathematically defined as \cite{levine2009, McCoy97}
\begin{align}
F_N:=\langle E_N\rangle-T\langle S_N\rangle,
\label{eq_fhe}
\end{align}
where $\langle E_N\rangle=\langle E^{id}_N\rangle+\langle U_N\rangle$ is the sum of the kinetic and potential energy, and -$T\langle S_N\rangle$ is the mean heat exchanged between the system and its environment. Replacing $\langle E_N\rangle$ and $T\langle S_N\rangle$ with their equivalent expressions in Eq.(\ref{eq_fhe}) and using Eq.($\ref{eq_2bodynew}$), the following expression for a closed $NVT$ system at high (near-freezing) or near-zero densities is obtained:
\begin{align}
F_N=F_N^{id}(T)+T\left[\frac{Nk_B}{2}-\delta S_N^{(2)}\right],
\label{eq_fheperparticle}
\end{align}
where $F_N^{id}(T)$ is the ideal-gas Helmholtz free energy.

The system relaxation time, denoted $\tau_{sys}$, is defined as the inverse rate of approach to (thermodynamic) equilibrium. Based on Eq.($\ref{eq_fheperparticle}$), the Helmholtz free energy equilibrates once the temperature $T$ and residual two-body excess entropy reach equilibrium. Thus, for such systems
\begin{align}
\tau_{sys}=max(\tau_T,\tau_{\delta S}),
\label{eq_relaxsystem}
\end{align}
where $\tau_{T}$ and $\tau_{\delta S}$ are the relaxation times for the temperature $T$ and residual two-body excess entropy, respectively. In statistical mechanics, the (kinetic) temperature $T$ can be defined via the equipartition theorem \cite{Waterson51} as the mean kinetic energy per atom ($T:=2\langle E^{id}_N\rangle/3Nk_B$). It is, therefore, a \textit{local} quantity which should, in general, equilibrate \textit{faster} than the residual two-body entropy, whose equilibration depends on the equilibration of all the particles' positions within the system, (at least for local and particle-wise thermostating). Thus, we typically always have $\tau_{\delta S}>\tau_{T}$ and
\begin{align}
\tau_{sys}=\tau_{\delta S}.
\label{eq_finalrelaxsystem1}
\end{align}

In nonequilibrium thermodynamics \cite{de1984non}, the definition of entropy, i.e., Eq.(\ref{eq_entropy}), is also used for systems near equilibrium. However, this assumption may not be valid for systems in which the local equilibration hypothesis is \textit{never} satisfied, such as in glasses \cite{Vilar11081,rubi97}. For nonglassy (atomic) systems in contact with a local thermostat, such as the Langevin or dissipative particle dynamics (DPD) thermostat, one would expect the temperature $T$ is equilibrated before local equilibration is reached (for systems with local thermostats, in local equilibrium, temperature is in global equilibrium, whereas structure is in local equilibrium). Hence, for such systems, Eq.(\ref{eq_2bodynew1}) is generalized for
$t>\tau_T$, ensuring the systems have most likely reached local equilibrium, as
\begin{align}
\delta S_N^{(2)}(t)=\frac{N\rho}{2T}\int d\mathbf{r}g_N(r,t)\delta F_N(r,t),
\label{eq_entropyevoloution1}
\end{align}
with
\begin{align}
\begin{split}
   g_N(r,t)&=e^{-w_N(r,t)/k_BT}\\
        &=e^{-u(r)/k_BT}e^{-\delta F_N(r,t)/k_BT}, 
\end{split}
\label{eq_rdfevolnew}
\end{align}
where the second line comes from generalizing Eq.(\ref{eq_pmf2}).
Equation (\ref{eq_entropyevoloution1}) implies that $\tau_{\delta S}$ is equal to $\tau_{\delta F}$, the relaxation time for $\delta F_N(r,t)$. Thus, for nonglassy atomic systems at near-freezing or zero densities, $\tau_{sys}=\tau_{\delta F}$ [refer to Eq.(\ref{eq_finalrelaxsystem1})].

In generic physical systems, it may not be always feasible to model the relaxation processes of dynamical quantities with an exponentially decaying function, especially when those systems undergo at least one phase transition. Nonetheless, one would expect that most of the effects of phase transitions on relaxation processes disappear once local equilibrium is reached. If $\delta F_N(r,t)$ in Eq.(\ref{eq_rdfevolnew}) is substituted with (the exponential ergodicity hypothesis \cite{Davidchack2009})
\begin{align}
\delta F_N(r,t)=\delta F_N(r)+\delta F^{\prime}_{N}(r)e^{-t/\tau_{\delta F}}
\label{eq_ergodicity}
\end{align}
(for $t>\tau_T$), where the second term on the RHS indicates the extra (positive or negative) adiabatic work needed in nonequilibrium conditions to bring the two particles from infinity to a distance $r$, the instantaneous RDF for a nonglassy system (at \textit{any} density) is written as follows:
\begin{multline}
    \lim_{\:t\:>\:\tau_{T}}g_N(r,t)=g_N(r)e^{\alpha_N(r)e^{-t/\tau_{\delta F}}}\\=g_N(r)\{1+\alpha_N(r)e^{-t/\tau_{\delta F}}+O(e^{-2t/\tau_{\delta F}})\},
\label{eq_entropyevoloution}
\end{multline}
where
\begin{align*}
\alpha_N(r):=-\frac{\delta F_N^{\prime}(r)}{k_BT}
\end{align*}
is a signed dimensionless physical quantity including nonequilibrium (structural) information. In the above Taylor series, the higher-order terms are neglected as they decay much faster than the first two terms in the brackets.  This paper is only concerned with situations in which $N>>1$. Hereafter, we drop the subscript $N$ from all the quantities to stress finite-size effects are negligible for $N>>1$.

In computer simulations, if the initial structure is more \textit{expanded} compared to the final structure, one would expect $\delta F^{\prime}$ to be, on average, positive (equivalent to $\alpha<0$); This is because the initial number density distribution is more uniform than it should be. Consequently, \textit{less} spots are, on average, available to bring two atoms from infinity to a distance $r$ without disrupting the structural stability. Hence, one needs some extra \textit{positive} work to make space for them. On the other hand, if the initial structure is more \textit{compacted} than the final structure, one would expect $\delta F^{\prime}$ to be, on average, negative (equivalent to $\alpha>0$); This is because there will be \textit{more} marginal space initially available (in the simulation box) than in the final state, which makes moving two atoms from infinity to a finite distance without changing the position of other atoms more likely, leading to averagely less total (adiabatic) work or $\delta F^{\prime}<0$ [refer to Eq.(\ref{eq_ergodicity}); $\delta F(r,t)<\delta F(r)\Rightarrow\delta F^{\prime}<0$].

Equation (\ref{eq_entropyevoloution}) can be further simplified to (at any density)
\begin{equation}
    \lim_{\:t\:>\:\tau_T}\lim_{\:r\:>\:\xi}g(r,t)=1+\alpha e^{-t/\tau_{\delta F}}
\label{eq_tcfevolution12}
\end{equation}
since for $r> \xi$, the RDF $g(r)\approx1$ and $\alpha(r)$ is expected to be (effectively) $r$ independent, provided the structural effects of the system-thermostat coupling are small enough in order not to significantly perturb the system's natural dynamics.
At near-zero densities, $\xi\approx r_c$ and $\tau_{\delta F}\approx\tau_{sys}$. Substituting these values for near-zero densities in Eq.(\ref{eq_tcfevolution12}) gives
\begin{align}
 \lim_{\:t\:>\:\tau_T}\lim_{\:r\:>\:r_c}g(r,t)=1+\alpha e^{-t/\tau_{sys}}.
\label{eq_tcfevolution1234}
\end{align}
The sign of $\alpha$ determines how the RDF (during equilibration) decays to one in time. Equation (\ref{eq_tcfevolution1234}) is of particular significance since it can be used to study (possible) effects of initial structural inhomogeneity on  $\alpha$ and the time for the system relaxation process $\tau_{sys}$ in the Langevin model. In principle, Eq.(\ref{eq_tcfevolution1234}) can also be used at high densities (near the freezing point) for $r>\xi\:(\neq r_c)$. In practice, $\tau_{sys}$ decreases with increasing number density $\rho$ at a constant temperature \cite{Ben2009}. Hence, it is more difficult to observe the (long-range) evolution of the RDF in time at high densities compared to near-zero ones. We mainly focus on dilute systems in this paper.

\section{\label{methodology} Methodology and Simulation Details}

This article deals primarily with how the initial structure can affect the system relaxation time in the Langevin model. For this purpose, we performed many molecular-dynamics (MD) simulations utilizing the LAMMPS software package \cite{Plimpton1995} and computed the time-averaged RDF, $g(r)$, after the system's temperature $T$ is relaxed for three different well-known and well-used initial structures. The pair potential energy $u(r)$ is a truncated and shifted 6-12 Lennard-Jones (LJ) potential:
\begin{align}
  u(r) =
  \begin{cases}
                                V_{LJ}(r)-V_{LJ}(r_c), & \text{for $0< r\leq r_c$} \\
                                   0, & \text{for $r> r_c$} 
  \end{cases},
  \label{eq_tsLJ}
\end{align}
where $r_c=2.5\sigma$ is the cut-off radius, and
\begin{align*}
V_{LJ}(r)=4\varepsilon\:\left[\left(\frac{\sigma}{r}\right)^{12}-\left(\frac{\sigma}{r}\right)^6\right],
\label{eq_LJ}
\end{align*}
with $\sigma$ and $\varepsilon$  being the effective atomic radius and the dispersion energy (depth of the potential well), respectively. The 6-12-LJ equation approximates well the interactions in the noble massive gases (e.g., Ar and Kr), whose interactions are dominated by van der Waals forces. Note that the constant term $V_{LJ}(r_c)$ in the above equation is added to avoid any discontinuity at $r=r_c$ in the potential, leading to an impulsive contribution to the (thermodynamic) pressure \cite{frenkel2002}.

In a computer experiment, it is often convenient to express physical quantities in units other than the SI units. For example, by expressing these quantities in LJ reduced units allows one to benefit from the corresponding-states principle \cite{Tester97}. For the LJ reduced units, $k_B=m=\sigma=\varepsilon=1$. Also, using LJ reduced units reduces the floating-point/round-off error (caused by the limited precision of computational processors) as parameters are of order one \cite{frenkel2002}. We note that all the physical quantities quoted below are in LJ units. 

A pure monatomic fluid at number density $\rho_0=3.95\times10^{-2}=\rho_{cr}^{Ar}/8<<1$ ($\rho_{cr}^{Ar}$ is the critical density of Argon) was simulated in the Langevin canonical ensemble with zero total linear momentum and  with periodic boundaries to remove surface effects. In Refs. \cite{Roy99,White2008}, it is shown that in MD experiments with zero total linear momentum and periodic boundaries, the infinitesimal Galilean boost is conserved, which ensures each atom remains an inertia reference frame. In the Langevin model, the equation of motion for the $i$-th particle relative to the system center of mass reference frame ($R_{c.m.}=0$) is given by \cite{BUSSI2008}
\begin{equation}
\ddot{\mathbf{r}}_i(t)=\mathbf{F}_i+\mathbf{g}_i(t)\quad(i=1,2,\dots,N),
\label{eq_hamilton}
\end{equation}
where $\mathbf{F}_i$ is the net conservative force on the $i$-th particle, and $\textbf{g}_i(t)$ is a fictitious force, which modifies the dynamics of the system to account for the presence of the Langevin thermostat. $\textbf{g}_i(t)$ is responsible for the variations leading to the system thermalization and is given by \cite{BUSSI2008}
\begin{align}
\textbf{g}_i(t)dt=-\gamma\dot{\textbf{r}}_i(t)dt+\sqrt{2\gamma T}d\textbf{W}_i(t), 
\label{eq_correction1}
\end{align}
where $\gamma>0$ is the Langevin coupling constant, and $d\textbf{W}_i$ is a vector of independent Wiener processes, satisfying
\begin{equation}
\begin{gathered}
\langle dW_{ij}(t)dW_{kl}(t^{\prime})\rangle=\delta_{ik}\delta_{jl}\delta (t-t^{\prime})dt\\\langle dW_{ij}(t)\rangle=0\\(j,l=x,y,z) \text{ and } (i,k=1,\dots,N),
\end{gathered}
\label{eq_winernoise2}
\end{equation}
where $\delta (t-t^{\prime})$ is the Dirac delta distribution, $\delta_{ik/jl}$ is the Kronecker delta function, and $\langle\dots\rangle$ indicates an average over an ensemble of simultaneous, independent, and similar experiments.

In our computer experiments, the equations of motion (\ref{eq_hamilton}) are integrated numerically utilizing the GJF-2GJ algorithm, which is quite accurate, especially for the kinetic sampling of the phase space \cite{Jensen2019}. It has also been proven that the GJF-2GJ algorithm is capable of providing exact thermodynamic responses for constant and harmonic potentials for any timestep size $\delta t$ within the verlet stability criteria. In the GJF-2GJ algorithm, atom's velocities are computed at middle points of time steps. The $i$-th atom's half-step velocity relative to the system's c.m. ($R_{c.m.}=0$) is then given by \cite{Jensen2019_2}
\begin{align}
\textbf{v}_i(t+\frac{ \delta t}{2})=\frac{\textbf{r}_i(t+\delta t)-\textbf{r}_i(t)}{\sqrt{b}\delta t}\quad(i=1,2,\dots,N),
\label{eq_GJF2GJvelocity}
\end{align}
where $\textbf{r}_i$ is the on-site atomic position, $\delta t$ is the timestep size, and 
\begin{align}
b=(1+\frac{\gamma \delta t}{2})^{-1}
\label{eq_b}
\end{align}
to preserve semi-symplecticity. In computer experiments, the resulting equilibrium distributions are different from the true theoretical distributions as a result of discretization or truncation errors \cite{Mannella2006,Batrouni85}. To avoid losing physical information due to such systematic errors, the timestep should be much smaller than the inverse of the fastest vibrational frequency in the system. In this paper, we chose $\delta t = 0.0007$. Note that in numerical calculations with computers, the roundoff error $\propto1/\delta t$ \cite{kutz2013}. 

\begin{figure}
     \begin{subfigure}[b]{0.3\textwidth}
         \centering
         \includegraphics[width=\textwidth]{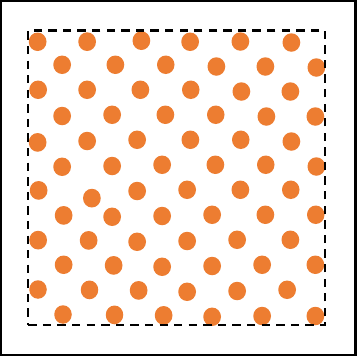}
         \caption{marginally inhomogeneous structure (FCC crystal)}
         \label{ch1a}
     \end{subfigure}
     \begin{subfigure}[b]{0.3\textwidth}
         \centering
         \includegraphics[width=\textwidth]{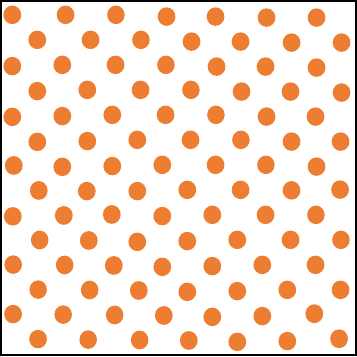}
         \caption{uniform structure (FCC crystal)}
         \label{ch1b}
     \end{subfigure}
     \begin{subfigure}[b]{0.3\textwidth}
         \centering
         \includegraphics[width=\textwidth]{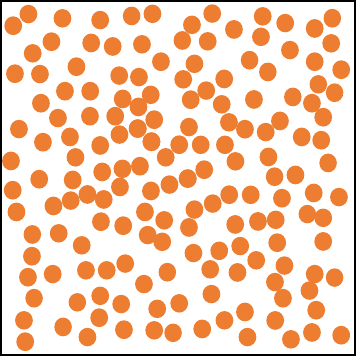}
         \caption{uniform structure (amorphous solid)}
         \label{ch_amorphous}
     \end{subfigure}
     \captionsetup{justification=raggedright,singlelinecheck=false}
        \caption{2D illustration of the initial structures.}
        \label{fig1}
\end{figure}

We first created $N_1\sim10^5$ LJ particles at temperature $T=0$ in the form of a face-centered cubic crystal, consisting of $43^3$ cells with lattice constant $a =(4/\rho_0)^{1/3}=4.66$ and locating at the center of a simulation box with a volume of $V=(45a)^3$. This structure is slightly marginally inhomogeneous in the number density of atoms [see Fig. \ref{ch1a}] due to the empty gap around the system's edges. Then, we created $N_2\sim10^5$ LJ particles under the same conditions, except this structure now fills the entire simulation box. Such a structure is macroscopically homogeneous in the number density of particles [see Fig. \ref{ch1b}]. Both of these initial structures are commonly used in MD simulations.  The second for the reason that macroscopically homogeneous densities are expected to be desirable initial conditions, and the first as a common approximation to the second without having to worry about lattice commensurability effects at the periodic boundaries. For Fig. \ref{ch1b}, LAMMPS is careful to put only one particle at the boundaries to avoid any unwanted atom overlap (using the \textit{box} style of the \textit{create\textunderscore atoms} command in LAMMPS).

Unless otherwise stated, for each case, the phase-space trajectories were first run for $t_e=\Delta t=175$ to equilibrate the system's temperature at a supercritical temperature $T=3.936$ for a few multiples of a basic Langevin thermal coupling constant of $\gamma=\gamma_0=5/7$. As will be confirmed in the next section, $t_e=175\:(>>\tau_T)$ is more than enough time to equilibrate the system's temperature for all the $\gamma$ values used.
The trajectories were then run for a duration of $t_{pro}=\Delta t=525$ so as to calculate the (time-averaged) RDF, $g(r)$, up to a maximum distance of $r_m=10 r_c=25$. Finally, to ensure that the results are not limited to only those systems with crystalline initial structures, we also performed a number of extra MD simulations with amorphous initial structures (randomly-distributed particles), which occupied the entire simulation box [see Fig. \ref{ch_amorphous}]. This is the only distinction between the second and third batches of simulations. It should also be noted that this kind of initial structure is typically (energetically) unstable because randomly-generated particles are often \textit{highly} overlapped. Therefore, we had to perform a \textit{local} pre-energy minimization using the Conjugate-Gradient (CG) algorithm for $\Delta t\simeq2.1$ for this initial state.

\subsection{Measurement of the time-averaged RDF }
In an MD experiment, the time-averaged RDF is measured as follows: First, a random atom is selected. Next, the algorithm computes $\delta n(r,t)$. Then, the algorithm increases the distance from $r$ to $r+\delta r$ and does exactly the same measurements \cite{Levine2011}. This procedure continues until $r$ reaches its maximum, that is $r_m=10\times r_c=25$. The algorithm repeats the whole process for another randomly-chosen atom, and so on. In the end, the algorithm determines the instantaneous RDF, defined as
\begin{align}
g(r,t):=\frac{1}{4\pi N \rho_0 r^2}\sum_{i=1}^N\frac{\delta n_i(r,t)}{\delta r},
\label{eq_meaurerdf}
\end{align}
where the summation is over particles and $\delta n_i(r,t)$ is the number of particles within the spherical shell of thickness $\delta r=25/256$ centered at the $i$-th atom between time $t$ and $t+\delta t$. The time-averaged RDF is computed by averaging $g(r,t)$ over time:
\begin{align}
g(r)=\frac{\delta t}{t_{pro}}\sum_{j=1}^{t_{pro}/\delta t}{g(r,t_e+j\delta t)}
\label{eq_TARDF}
\end{align}
with $t_e=175$ and $t_{pro}=525$ being the equilibration and production time, respectively. In the previous section, we derived an (approximate) expression for $g(r,t)$ for $r>r_c$ and $t>\tau_T$. By substituting Eq.(\ref{eq_tcfevolution12}) into Eq.(\ref{eq_TARDF}), one finds that if the system is not \textit{well} equilibrated, the time-averaged RDF undergoes a shift. For high (near the freezing point) and near-zero densities,
\begin{align}
\lim_{\:t_e\:>\:\tau_T}\lim_{\:r\:>\:\xi}[g(r)-1]= Ae^{-t_e/\tau_{sys}},
\label{eq_TARDF1}
\end{align}
where $\xi\approx r_c$ at low densities, and
\begin{equation*}
    A=-\frac{\delta F^{\prime}}{k_BT}\frac{\tau_{sys}}{t_{pro}}\left(1-e^{-t_{pro}/\tau_{sys}}\right)\propto-\delta F^{\prime}.
\end{equation*} 
In Langevin dynamics, Eq.(\ref{eq_TARDF1}) may be generalized as follows to incorporate effects of the system-thermostat coupling (weak coupling):
\begin{align}
\lim_{\:t_e\:>\:\tau_T(\gamma)}\lim_{\:r\:>\:\xi}[g(r)-1]= A(\gamma)e^{-t_e/\tau_{sys}(\gamma)}
\label{eq_TARDF2}
\end{align}
In this paper, we study Eq.(\ref{eq_TARDF2}) for some (common) values of $\gamma\;(\sim1)$ to see how the RHS may respond to the three well-used initial structures in a molecular-dynamics simulation.

\section{\label{result} Results and Discussion}
\subsection{Temperature relaxation time}

\begin{figure}
     \centering
     \begin{subfigure}[b]{0.48\textwidth}
         \centering
         \includegraphics[trim=1 1 1 1, clip,width=\textwidth]{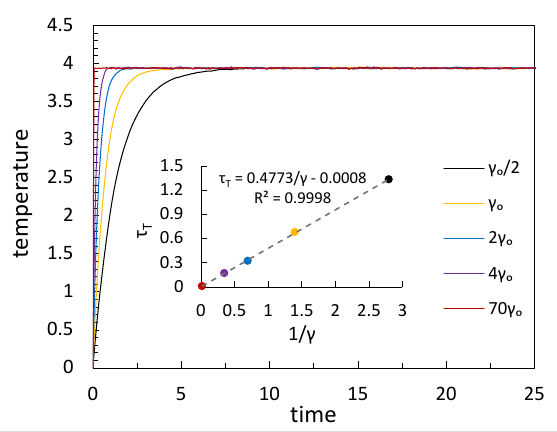}
         \caption{uniform initial structure}
         \label{ch2}
     \end{subfigure}
     \begin{subfigure}[b]{0.48\textwidth}
         \centering
         \includegraphics[trim=1 1 1 1, clip, width=\textwidth]{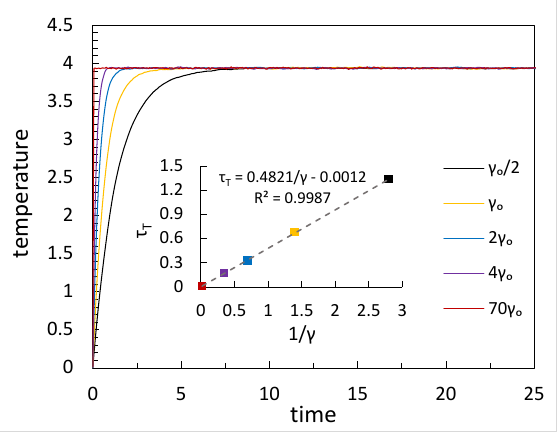}
         \caption{marginally inhomogeneous initial structure}
         \label{ch2b}
     \end{subfigure}
     \captionsetup{justification=raggedright,singlelinecheck=false}
        \caption{Temporal evolution of the instantaneous (kinetic) temperature $T$ during the equilibration phase for both kinds of initial structures and some multiples of $\gamma_0=5/7$. As can be observed, the higher the value of the Langevin coupling constant $\gamma$, the faster the system reaches thermal equilibrium. Both insets depict that the (kinetic) temperature relaxation time, $\tau_T$, varies linearly with $1/\gamma$. To be precise, $\tau_T\approx1/2\gamma$. (Note that all the values are in the LJ units.)}
        \label{fig2}
\end{figure}

In Fig. \ref{ch2}, we plotted the evolution of instantaneous temperature $T$ for the homogeneous crystalline structure shown in Fig. \ref{ch1b} for some multiples of a Langevin coupling constant, $\gamma_0=5/7=1/2000\delta t$, to demonstrate the system is in thermal equilibrium for all $t\geq175$. This figure shows the \textit{stronger} the temperature coupling between the system and the Langevin thermostat, the \textit{faster} the system reaches thermal equilibrium. As expected, for an exponentially-ergodic system, the time evolution of quantities, such as temperature $T$, can be approximated and well fitted by an exponential curve \cite{Basconi2013}. The relaxation time for quantity $X$, denoted by $\tau_X$, is then the (characteristic) time constant of the fitted exponential function. We have computed the temperature relaxation time for some $\gamma$ values and plotted them versus $(\gamma)^{-1}$ in the inset. As is expected, the temperature relaxation time $\tau_T$ varies \textit{linearly} with the inverse of the thermostat coupling constant $\gamma$ as $\tau_T\approx(\gamma)^{-1}/2$. In Langevin dynamics, by using the Ito's Lemma Chain Rule \cite{Gardiner2009}, we obtain the following expression for the system's kinetic temperature $T$ during equilibration \cite{BUSSI2008}:
\begin{align}
\langle T(t)\rangle=(T_i-T_f)e^{-t/\tau_T}+T_f,	
\label{eq_kinetic}
\end{align}
where $\tau_{T}=1/2\gamma$ is the temperature relaxation time, and $T_f$ ($T_i$) is the final (initial) equilibrium temperature. For this paper, $T_i=0$ and $T_f=3.936=3\times T_{cr}^{Ar}$ ($T_{cr}^{Ar}$ is the critical temperature of argon).

In Fig. \ref{ch2b}, similar to the homogeneous structure, we plotted the temporal evolution of temperature during the system's temperature equilibration for the same $\gamma$ values, but now for the  inhomogeneous structure shown in Fig. \ref{ch1a}. Then, we have calculated the temperature relaxation time for each $\gamma$ and plotted them versus $1/\gamma$ in the inset. As expected, the temperature relaxation time varies with the Langevin coupling constant $\gamma$ as $\tau_T\approx1/2\gamma$, which is in excellent agreement with Eq.($\ref{eq_kinetic}$). The simulations with the amorphous initial structure [see Fig. \ref{ch_amorphous}] also show the same outcomes (not displayed here). From the above figures, it is clear that $t_e=175>>\tau_T$ and the temperature relaxation time is independent of the initial structure in Langevin dynamics. 

\begin{figure}
     \centering
     \begin{subfigure}[b]{0.48\textwidth}
         \centering
         \includegraphics[trim=1 1 1 1, clip,width=\textwidth]{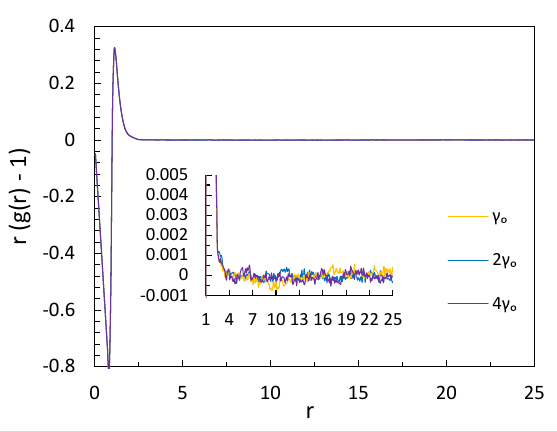}
         \caption{\centering uniform initial structure}
         \label{ch4}
     \end{subfigure}
     \begin{subfigure}[b]{0.48\textwidth}
         \centering
         \includegraphics[trim=1 1 1 1, clip, width=\textwidth]{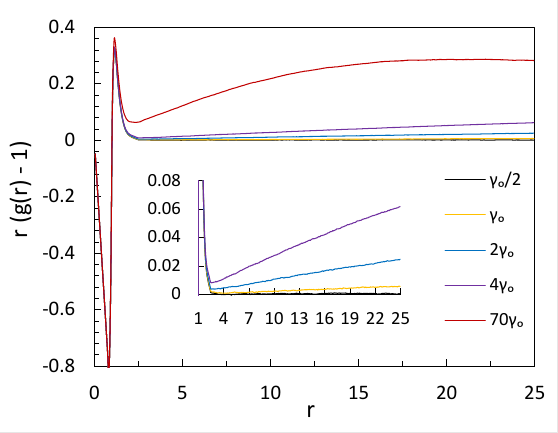}
         \caption{\centering marginally inhomogeneous initial structure}
         \label{ch3}
     \end{subfigure}
     \captionsetup{justification=raggedright,singlelinecheck=false}
        \caption{$r[g(r)-1]$ for both kinds of initial structures, $t_e= 175$, and some multiples of $\gamma_0=5/7$. These figures reveal that,  although for the uniform initial structure, $r[g(r)-1]$ is independent of $\gamma$ at thermal equilibrium, it is, indeed, highly $\gamma$ dependent for the marginally inhomogeneous initial structure. (Note that the insets are a zoomed-in version of the figures and all the values are in the LJ units.)}
        \label{fig3}
\end{figure}

\subsection{Examining Eq.(\ref{eq_TARDF2})}
For dilute systems and $r>r_c$ we expect $[g(r)-1]$ to be infinitesimally small if the system is fully equilibrated.  To make any deviation from the equilibrium more easily visible, in this section, we plot $r[g(r)-1]$, instead of $[g(r)-1]$ (i.e., we scale it by $r$). For Fig. \ref{ch4}, we plotted $r[g(r)-1]$ for the case of the uniform structure of Fig. \ref{ch1b}, the equilibration time $t_e=175>>\tau_T$ and some multiples of $\gamma=\gamma_0\approx0.7$. This figure demonstrates that the RDF is independent of $\gamma$ for $t_e=175$. Thus, $t_e>>\tau_{sys}$ for $\gamma=\gamma_0,\:2\gamma_0,\:4\gamma_0$ [based on Eq.(\ref{eq_TARDF2})]. At a very high value of $\gamma=70\gamma_0\:(=50)$, we found that the RDF is slightly out of equilibrium (not shown) which means the system relaxation time $\tau_{sys}$ has increased for $\gamma=70\gamma_0$ such that $t_e$ is not much longer than $\tau_{sys}$ anymore.  Afterwards, we repeated the simulations for the case of the randomly-dispersed structure of Fig. \ref{ch_amorphous}. As is expected, we found exactly the same behaviour for $r[g(r)-1]$ (not shown here). This would suggest that the system relaxation times of initial structures which are, on average, uniformly distributed throughout the simulation box are of the same order and only weakly dependent on the Langevin coupling constant (increasing behaviour).

\begin{figure}
     \centering
     \begin{subfigure}[b]{0.48\textwidth}
         \centering
         \includegraphics[trim=1 1 1 1, clip,width=\textwidth]{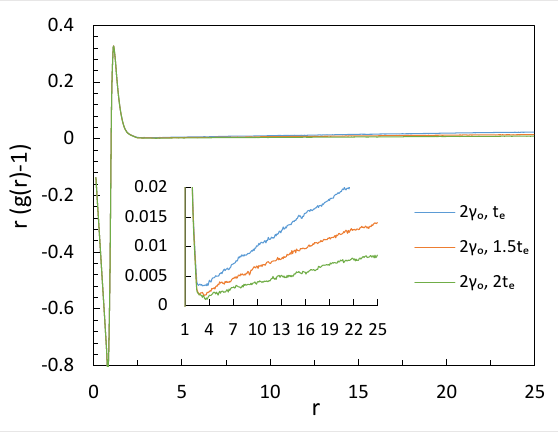}
         \caption{\centering nonuniform initial structure}
         \label{ch5a}
     \end{subfigure}
     \begin{subfigure}[b]{0.48\textwidth}
         \centering
         \includegraphics[trim=1 1 1 1, clip, width=\textwidth]{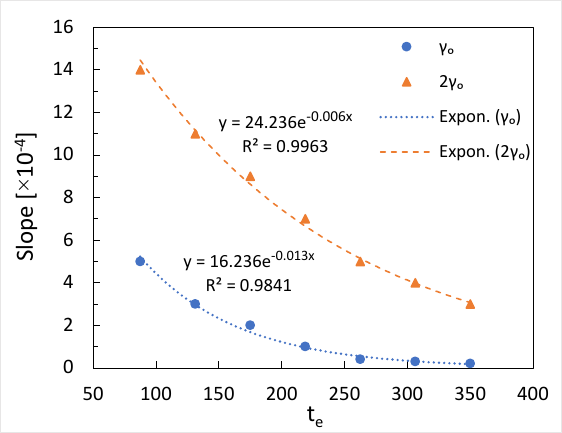}
         \caption{\centering nonuniform initial structure}
         \label{ch5b}
     \end{subfigure}
     \captionsetup{justification=raggedright,singlelinecheck=false}
        \caption{(a) Plot of $r[g(r)-1]$ versus $r$  for $\gamma=2\gamma_0=10/7$ and some multiples of $t_e=175$. As seen, the long-range slope decreases continuously with increasing the equilibration time.  (Note that the inset is a zoomed-in version of the figure and all the values are in the LJ units.) (b) The slope of  $\lim_{\:r\:>\:r_c}r[g(r)-1]$ vs. the equilibration time $t_e$ for $\gamma=\gamma_0=5/7$ and $2\gamma_0=10/7$. The dotted and dashed lines are the exponential fits to the data from our MD simulations. The time constant of each exponential fit is the system relaxation time [based on Eq.(\ref{eq_TARDF2})].}
        \label{fig4}
\end{figure}

\begin{figure}
    \centering
    \includegraphics[trim=1 1 1 1, clip, width=0.48\textwidth]{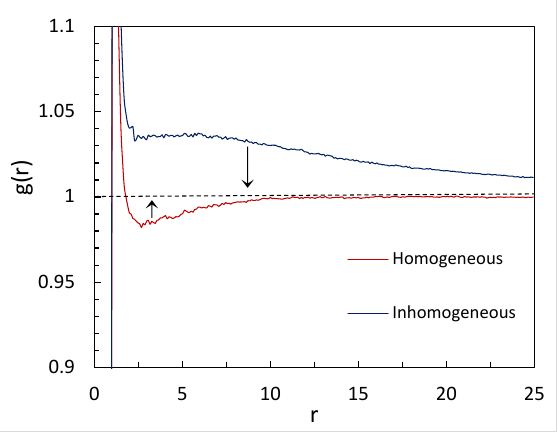}
    \captionsetup{justification=raggedright,singlelinecheck=false}
    \caption{How and in which direction the RDF evolves over time for two different initial structures at instant $t=2.8$. The RDF graph for the inhomogeneous case, unlike the homogeneous one, approaches one from above, which is the consequence of initial structural inhomogeneity. (Note that all the values are in the LJ units.)}
    \label{ch_gevolution}
\end{figure}

In Fig. \ref{ch3}, we plotted $r[g(r)-1]$ for the initial structure of Fig. \ref{ch1a} and the equilibration time $t_e=175\;(>>\tau_T$; the system is assumed to have already reached local thermal equilibrium), and some multiples of $\gamma_0=5/7$. As is expected from Eq.(\ref{eq_TARDF2}), $\lim_{\:r\:>\:r_c}r[g(r)-1]$ is $\gamma$ dependent in local equilibrium. In particular, $\lim_{\:r\:>\:r_c}r[g(r)-1]$ varies linearly with a $\gamma$ dependent slope for weak thermal couplings and nonlinearly for strong thermal couplings. Such deviation from zero is because (structural) entropy is not maximal and $\tau_T<<t_e<\tau_{sys}$. By comparing these two figures, we can conclude \textit{(hom: homogeneous, inhom: inhomogeneous)}
\begin{equation}
\tau_{sys}^{inhom}>>\tau_{sys}^{hom}.
\label{eq_finalcomparison}
\end{equation}
Therefore, an initial small marginal inhomogeneity in the Langevin model \textit{significantly} increases  the time required to equilibrate the system's structure. Further, increasing the Langevin constant (which reduces the thermal equilibration time) makes the equilibration significantly worse. The inhomogeneity in the marginal inhomogeneous initial state is primarily of a long-wavelength nature and the Langevin thermostat suppresses or slows down the evolution of the periodic and long-wavelength density fluctuations needed to equilibrate these modes.

In Fig. \ref{ch5a}, we plotted $r[g(r)-1]$ for the inhomogeneous structure for $\gamma=2\gamma_0=10/7$ and several multiples of $t_e=175$. This figure clearly reveals that the slope decreases with increasing the duration of the equilibration phase, i.e., $t_e$. Hence, the non-zero slope is basically because the amount of the equilibration time is \textit{not} enough to equilibrate the system's structure. Next, in Fig. \ref{ch5b}, we plotted the evolution of the slope with the equilibration time for $\gamma=\gamma_0,\;2\gamma_0$. We found that the slope varies with  $t_e$  in an exponential manner in local equilibrium, as exactly predicted in Eq.(\ref{eq_TARDF2}). In Fig. \ref{ch5b}, 
\begin{equation}
   \text{Slope}=\frac{\int_{r_c=2.5}^{r_{m}=25}[g(r)-1]dr}{r_{m}-r_c}=A(\gamma)e^{-t_e/\tau_{sys}(\gamma)}, 
\label{eq_slope}
\end{equation} 
where $A(\gamma)$ was found to be positive. The numerical results obtained from Fig. \ref{ch5b} for $\tau_{sys}(\gamma)$, i.e., $\tau_{sys}(\gamma_0)=(0.013)^{-1}$ and $\tau_{sys}(2\gamma_0)=(0.006)^{-1}\approx2\times\tau_{sys}(\gamma_0)$, suggest an increasing behaviour for the system relaxation time with increasing the Langevin constant $\gamma$.  This is in contrast to the (kinetic) temperature relaxation time $\tau_T$ which we saw decrease with increasing $\gamma$ [$\tau_{sys}(\gamma)\propto\gamma$ vs.  $\tau_T(\gamma)\propto1/\gamma$]. This increasing behaviour was, indeed, predictable
due to the structural disruptive effect caused by
the Langevin thermostat.

In Fig. \ref{ch_gevolution}, we plotted the RDF for $\gamma=\gamma_0$ at instant $t=2.8=4\tau_T(\gamma_0)$ for both homogeneous [see Fig. \ref{ch1b}] and inhomogeneous [see Fig. \ref{ch1a}] initial structures. As is seen, in the marginally inhomogeneous case, unlike the uniform one, $g(r)$ approaches one in time from \textit{above} over the system's relaxation. To be specific, based on Eq.(\ref{eq_TARDF2}), for the inhomogeneous structure (i.e., the compacted structure), $A>0$, and for the homogeneous case (the more expanded structure), $A<0$. This sign change was, indeed, predictable since $A\propto-\delta F^{\prime}$, the extra (positive/negative) adiabatic work to move two atoms during the structural equilibration from infinity to a distance $r$. In the Theory section, we showed that the sign of $\delta F^{\prime}$ depends on the initial structure, and in the inhomogeneous case, unlike the uniform one, it is negative ($\delta F_{inhom}^{\prime}<0$). This sign change can be considered as the second effect of initial structural inhomogeneity on the system structural relaxation. [$g(r,t)$ graphs evolve in the directions shown in Fig. \ref{ch_gevolution} before the system reaches equilibrium.] 

\begin{figure}
    \centering
    \includegraphics[trim=1 1 1 1, clip, width=0.48\textwidth]{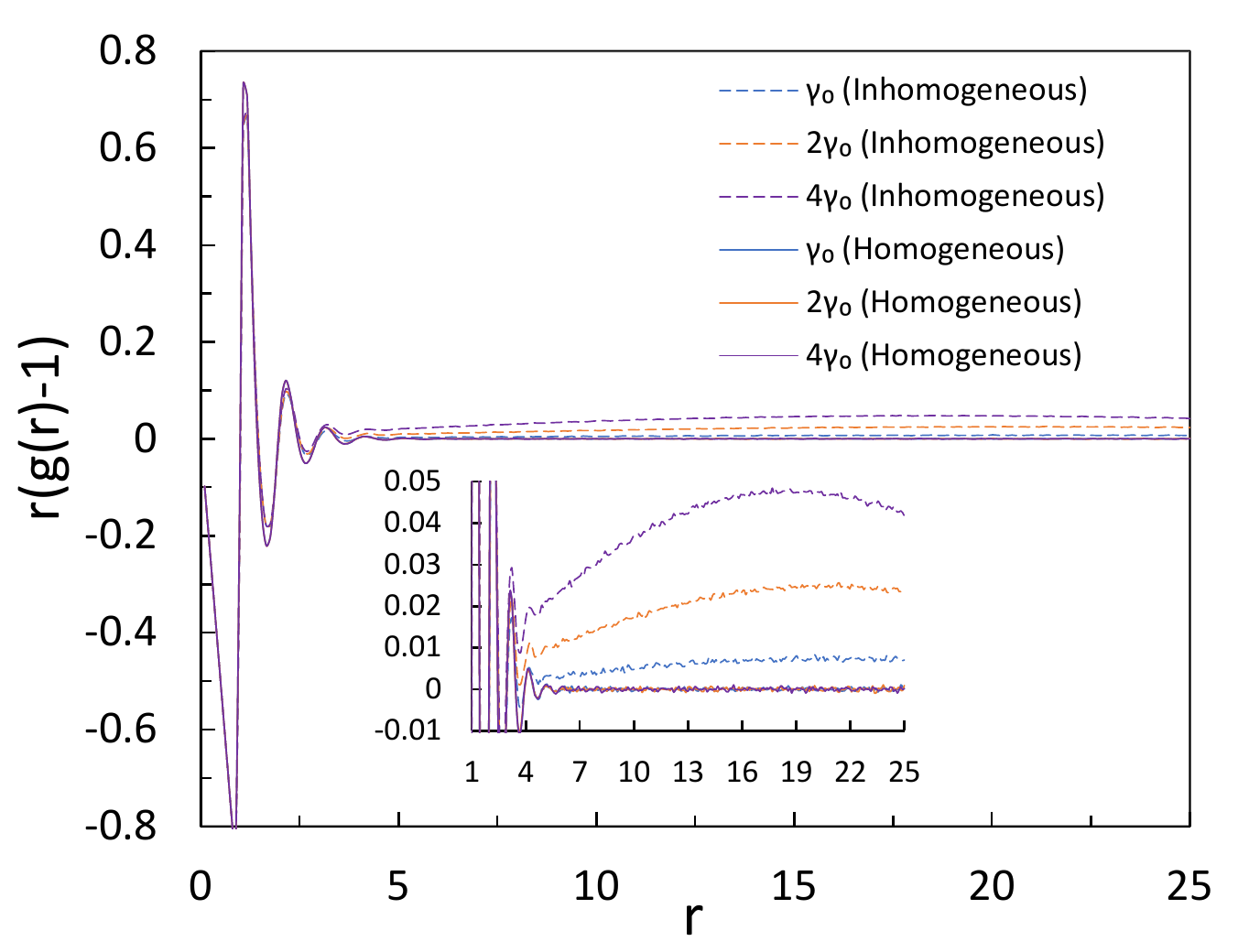}
    \captionsetup{justification = raggedright,singlelinecheck=false}
    \caption{$r[g(r)-1]$ at density $\rho=16\times\rho_0=0.632$ for both kinds of initial structures, $t_e= 7$, $t_{pro}=4.9$, and some multiples of $\gamma_0=5/7$. This figure demonstrates the RDF in all the homogeneous cases  is fully equilibrated regardless of the value of $\gamma$.  However, for the marginally inhomogeneous cases,  $r[g(r)-1]$ is $\gamma$ dependent, indicating the RDF has not been well equilibrated (Note that the inset is a zoomed-in version of the figure, and all the values are in the LJ units.)}
    \label{ch_high_density}
\end{figure}

Finally, to confirm that the low-density results are also applicable to higher densities, we have plotted $r[g(r)-1]$ in Fig. \ref{ch_high_density} at density $\rho=16\times\rho_0$ for homogeneous [Fig. \ref{ch1b}] and inhomogeneous [Fig. \ref{ch1a}] initial structures, $\gamma=\gamma_0,\:2\gamma_0,\:4\gamma_0$,  $t_e=7\geq10\tau_T$, and $t_{pro}=4.9\geq7\tau_T$. As expected, the homogeneous graphs, unlike the inhomogeneous ones, are \textit{all} fully equilibrated for the same values of $t_e$ and $t_{pro}$. Hence, one would conclude, assuming exponential ergodicity, $\tau_{sys}^{inhom}>\tau_{sys}^{hom}$ (and $\uparrow\tau_{sys}\sim\;\uparrow\gamma$) also holds at higher densities. An interesting point about this figure is that $\lim_{r>\xi}A$ for $\gamma=4\gamma_0=2.86$ (similar to the case of $\gamma=70\gamma_0=50$ at $\rho=\rho_0$) is $r$ dependent for $r>\xi$; this is because this system is now in the Brownian dynamics regime, that is strong thermal couplings. These observations imply that at higher densities, the minimum $\gamma$ value required for Brownian dynamics is smaller than that for low densities.
Figure \ref{ch_high_density} demonstrates the validity of our theory and its predictions at high densities.

\section{\label{conclusion} Conclusion}

In this paper, we found that the system relaxation time is not always independent of the initial state in molecular-dynamics simulations. In particular, in Langevin dynamics, for a marginally inhomogeneous initial structure compared to macroscopically homogeneous initial structures, the rate at which the structure equilibrates is much lower; This is probably because imposing marginal inhomogeneity in a system under periodic boundary conditions introduces long-wavelength inhomogeneous modes whose fluctuations are suppressed by the Langevin thermostat thereby slowing equilibration. While these results might seem intuitive, we have placed them on a solid analytical foundation in this paper.  In addition, we found initial structural inhomogeneity makes the RDF approach one in time from \textit{above} at large distances. These effects of initial structural inhomogeneity in the Langevin $NVT$ ensemble should be taken into account while investigating the structural evolution of systems at near-zero and high densities in (MD) simulations. As DPD and Langevin thermostats are similar \cite{Pastorini2007}, we would expect almost the same behaviours for the DPD thermostat.

This paper showed that the structural relaxation time, unlike the temperature relaxation time $\tau_T$, is an increasing function of the Langevin constant, $\gamma$, in Langevin dynamics. In (MD) simulations, to ensure the equilibration time $t_e$ is enough for the system to equilibrate at low/high densities, one should plot $r[g(r)-1]$ for $r>\xi$ [$\xi$ denotes the (effective) correlation length]. Any state- or thermostat-dependent (non)linear behaviour in $r[g(r)-1]$ indicates the system has not equilibrated properly and hence some nonequilibrium values might have leaked into the equilibrium time averaging of dynamical quantities, which leads to unreliable averaged equilibrium values.

The data that supports the findings of this study are available within the article in the figures.

\section{Acknowledgements}

This work was supported by the Natural Sciences and Engineering Council of Canada (NSERC). We would like to thank the Shared Hierarchical Academic Research Computing Network (SHARCNET) and Compute/Calcul Canada for the computational resources.

\bibliography{MozafarDenniston}

\end{document}